\documentclass[twocolumn,pra,superscriptaddress]{revtex4}
\usepackage{amsmath,amssymb,}
\usepackage{mathrsfs}
\usepackage{graphicx}
\usepackage{dcolumn}
\usepackage{bm}

\usepackage[colorlinks,citecolor=red]{hyperref}

\usepackage{rotating}
\usepackage{floatrow}
\usepackage{braket}
\usepackage{lipsum}

\usepackage{bbm}

\newcommand{\Rmnum}[1]{\expandafter\@slowromancap\romannumeral #1@}
\newcommand*{\hatH}{\hat{H}}

\begin{document}

\title{Realizing and detecting Stiefel-Whitney insulators in an optical Raman lattice}

\author{Jian-Te Wang}
\email{dz1922027@smail.nju.edu.cn}

\author{Jing-Xin Liu}

\author{Hai-Tao Ding}
\affiliation{National Laboratory of Solid State Microstructures, School of Physics,
and Collaborative Innovation Center of Advanced Microstructures, Nanjing University, Nanjing 210093, China}
\affiliation { Key Laboratory of Atomic and Subatomic Structure and Quantum Control (Ministry of Education), Guangdong Basic Research Center of Excellence for Structure and Fundamental Interactions of Matter, School of Physics, South China Normal University, Guangzhou 510006, China}

\author{Peng He}
\email{penghe@hku.hk}
\affiliation{Department of Physics, Guangdong-Hong Kong Joint Laboratory of Quantum Matter, The University of Hong Kong, Pokfulam Road, Hong Kong, China}




\begin{abstract}
We propose a feasible scheme to realize a four-band Stiefel-Whitney insultor (SWI) with spin-orbit coupled ultracold atoms in an optical Raman lattice. Four selected spin states are coupled by carefully designed Raman lasers, to generate the desired spin-orbit interactions with spacetime inversion symmetry. We map out a phase diagram with respect to the experimental parameters, where a large topological phase region exists. We further present two distinct detection methods to resolve the non-abelian band topology, in both equilibrium and dynamical ways. The detection relies on the spin textures extracted from the time-of-flight imaging, showing the tomographic signatures in the ground states and long-time averaged patterns on certain submanifold via a bulk-surface duality. Our work paves a realistic way to explore novel real topology with quantum matters.

\end{abstract}

\maketitle

\section{Introduction}

The study of topological insulators (TIs) has been a major focus in  ultra-cold atoms \cite{DWZhang2018,Goldman2014a,FMei2012_PRA,DWZhang2016_PRA} and condensed matter physics \cite{Hasan2010,XLQi2011,CKChiu2016,XGWen2017} in the past decades. The early tenfold classification based on fundamental symmetries, including anti-unitary time-reversal $\mathcal{T}$, particle-hole $\mathcal{C}$, and chiral $\mathcal{S}$ symmetry in noninteracting fermionic systems, highlights the central role of the system symmetries \cite{Schnyder2008,Kitaev2009,Ryu2010}. This approach has been extended to unitary spatial symmetries \cite{Kruthoff2017,Po2017}, leading to the theoretical discovery of crystalline symmetry protected topological phases, such as the fragile topological insulators \cite{Po2018,Bradlyn2019,Kooi2019,Bouhon2019,Slager20202} and the higher-order topological insulators \cite{Benalcazar2017,Benalcazar20172,Schindler2018,BXie2021}. In particular, the topological Euler phase (TEP) and Stiefel-Whitney insulator (SWI) protected by the combined symmetry of parity-time ($\mathcal{PT}$) with spatial inversion ($\mathcal{P}$) or the two-fold rotations and TRS ($\mathcal{C}_2\mathcal{T}$), has attracted considerable interests \cite{CFang2015,YXZhao2017,YXZhao2020,Slager20201,Slager20203,JAhn2018,JAhn2019,QWu2019,Ezawa2021,Takahashi2023}. The SWI is characterized by the second Stiefel-Whitney (SW) class of the real ground states imposed by the protecting symmetry, which is a crystalline-protected analogue of the Chern number. The classification based on orthogonal K-theory for the single-gap topology has been established \cite{YXZhao2016}, and the non-Abelian aspect of multi-gap topology has been widely explored \cite{QWu2019,Slager20201}. Meanwhile, the SWI also manifests its intriguing features in the semimetallic phases, such as the existence of the point nodes and line nodes carrying a $\mathbb{Z}_2$ monopole charge \cite{CFang2015,YXZhao2017,JAhn2018}, and higher dimensional generalizations \cite{Lim2023,Bouhon2023}. Therefore, it is of paramount importance to verify the theoritcal findings in a feasible experimental setup.

Ultracold atoms \cite{DWZhang2018,Goldman2014a} provide a versatile platform for quantum simulation due to the high tunability of the atom-light interactions \cite{WZhao2022,SLZhu2007,LBShao2008,Tarruell2012,Jotzu2014,BSong2019,Minguzzi2022,DWZhang2020_SC,SLZhu2013_PRL,DWZhang2020_PRB}. The creation of artificial gauge fields \cite{Dalibard2011,SLZhu2006,Beeler2013,YJLin2009,Ruseckas2005,Jaksch2003,Goldman2009,Gorg2019} and spin-orbit coupling (SOC) \cite{YJLin2011,Galitski2013,XFZhou2013,XCui2014,BZWang2018,JHZhang2022,GLiu2010_PRA,DWZhang2012_PRA,SLZhu2011_PRL,ZCXu2022} has led to a plethora of experimental demonstration of topological phases \cite{LHuang2016,ZWu2016,ZYWang2021,JZLi2022_PRL,QXLv2021_PRL, QXLv2023_PRA}. The fast development of detection approaches based on dynamical response \cite{Abanin2013,Atala2013,Grusdt2014,Aidelsburger2015,Dauphin2013} and band tomography \cite{Alba2011,Hauke2014,Li2016,RBLiu2015} has also contributed greatly to experimental progress. However, previous experiments usually only involve abelian bands in an optical lattice. The engineering and detection of topological bands with degeneracy and stabilized by certain symmetries are still challenging.

In this paper, we propose a practical approach to realizing a two-dimensional (2D) $\mathcal{PT}$-symmetric SWI in an optical Raman lattice. We include four selected spin states in the ground state manifold so that transitions driven by the same Raman potentials share identical Clebsch-Gorden coefficients. This results in the desired real hopping events, which give rise to a low-energy s-band model that naturally preserves the $\mathcal{PT}$-symmetry. The two degenerate s-bands are indexed by the second SW class. We map out a phase diagram according to the second SW class and find a robust region of SWI on the phase plane. Since the total Chern number is zero and the degeneracy for occupied bands in SWI doesn't possess any featured points in momentum space, detection through Hall drift \cite{Aidelsburger2015,Dauphin2013} and Ramsey interferometry \cite{Abanin2013,Atala2013,Grusdt2014} is not suitable. To overcome this obstacle, we further show the robustness of the SWI in our model by introducing a $\mathcal{T}$-broken term modified under an external magnetic field. We also provide two detection methods to resolve the non-Abelian band topology, based on equilibrium and dynamical schemes. The equilibrium scheme requires tomography of the prepared degenerate ground state through the quasi-momentum distribution of the spin textures extracted from time-of-flight (TOF) imaging. To circumvent the O(2) gauge mixing of the raw data from state tomography, we apply a parallel transport gauge and calculate the second SW class, giving a direct probe of the topology. The dynamical scheme probes the unitary evolution following a sudden quench. The topological information is rebuilt by the long-time averaged spin textures on the reduced quasi-momentum submanifold called the band inversion surface (BIS), which is defined by the quenching axis. We show the bulk-surface duality for our model by generalizing the original $\mathbb{Z}$ class case~\cite{LZhang2018,XLYu2021}.

The rest of this paper is organized as follows. In Sec. \ref{model}, we present a Raman scheme coupling four selected internal states for alkali atoms to realize a 2D SWI in a square optical lattice. To reveal nontrivial topology in our proposed model, two dintinct detection methods are discussed, with an equilibrium scheme shown in Sec. \ref{detect_eq} and a dynamical one shown in Sec. \ref{detect_dyn}, based on achievable experimental techniques.   

\section{The model}\label{model}

\begin{figure}[htbp]
	\centering
	\includegraphics[width=\textwidth]{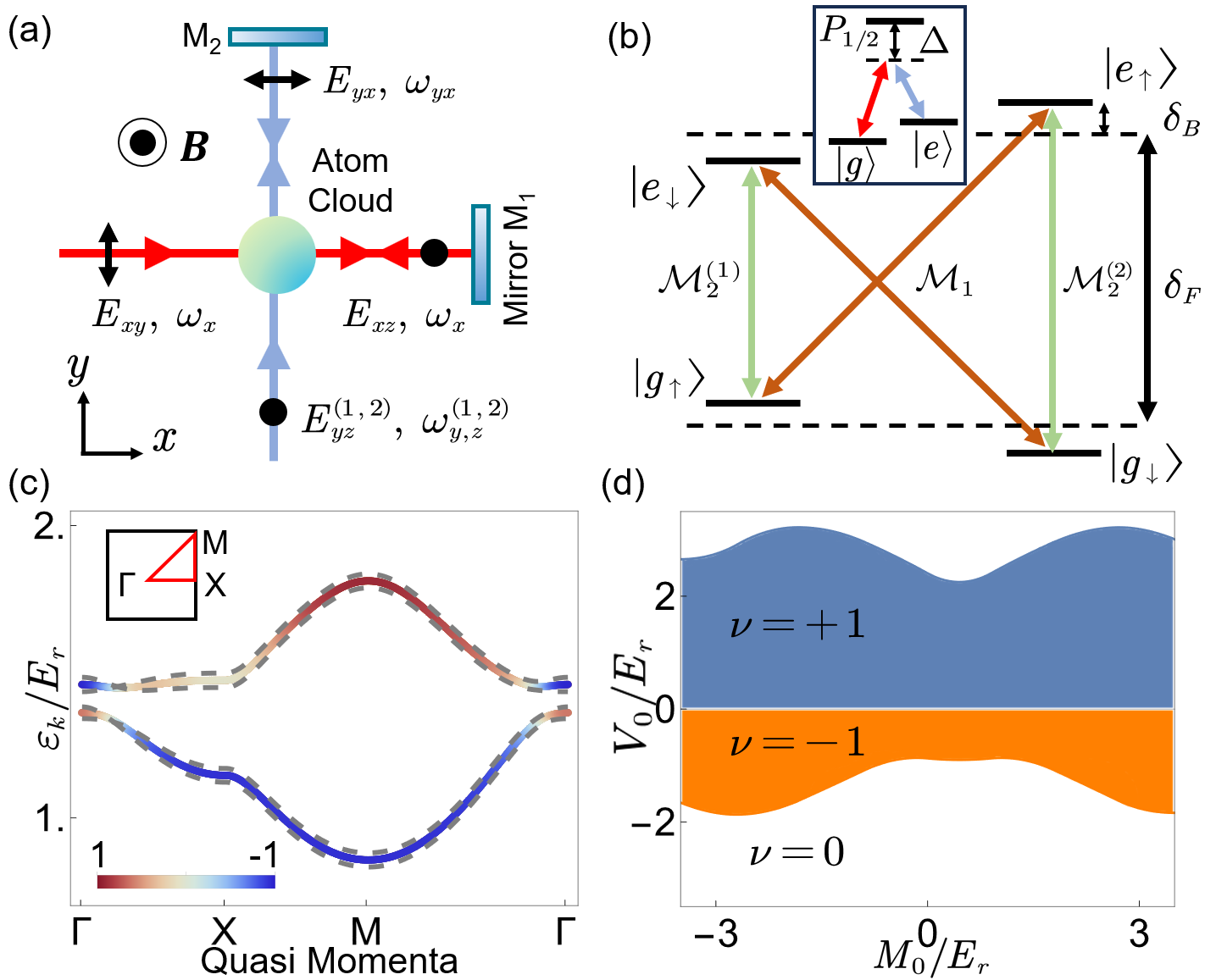}
 \caption{(a) Schematic of the setup for realizing the Stiefel-Whitney insulator (SWI). The cold atom cloud is confined by a magnetic trap and illuminated by laser fields, $\pmb{E}_x$ and $\pmb{E}_y$, reflected by mirrors, which create 2D optical lattice potentials and Raman couplings in the $x-y$ plane. (b) Related Raman transitions between the involved spin states. Raman transitions $\mathcal{M}_1^{(1,2)}$ and $\mathcal{M}_2$ are driven by distinctive two-photon processes with individual polarization and frequency configurations. All these two-photon processes occur between $\ket{e}$ and $\ket{g}$ states, as shown in the inset. (c) The s-band structures along the high-symmetry lines in the first Brillouin zone (shown in the inset) for the $\mathcal{T}^2=-1$ symmetric case (solid lines, $m_z=0$) and the broken case (dashed lines, $m_z=0.05E_r$). The color indicates the value of $\braket{\gamma_3}$ for the corresponding eigenstates. The global degeneracy is lifted in the broken case. Parameters are chosen as $V_0=3E_r$, $M_{10}=M_{20}=M_0=E_r$, and $\delta_V=0.3V_0$. (d) The second SW class of the lowest bands with respect to the lattice depth $V_0$ and the Raman coupling strength $M_0$. $E_r=(\hbar k_L)/2m_a$ is the recoil energy, with $m_a$ being the mass of the atom.}\label{fig1}
\end{figure}

We start by considering the realization scheme with ultracold alkali atoms in an 2D tunable optical Raman lattice, as illustated in Fig. \ref{fig1}(a) and \ref{fig1}(b). To implement the real degenerate bands, we use four Zeeman-split ground hyperfine levels in ground state manifold $S_{1/2}$, namely $\ket{e_{\uparrow,\downarrow}}=\ket{F+1,\mathrm{m}F=1,-1}$ and $\ket{g_{\uparrow,\downarrow}}=\ket{F,\mathrm{m}F=-1,1}$, with energy landscape shown in Fig. \ref{fig1}(b). The Raman coupling potentials are generated by a monochrome light field $\pmb{E}_x$ and a multifrequency beam $\pmb{E}_y$: $\pmb{E}_x=E_{xz}\pmb{e}_z\cos{k_Lx}+\mathrm{i}E_{xy}\pmb{e}_y\sin{k_Lx}$, and $\pmb{E}_y=\sum_{i=1,2}E_{yz}^{(i)}\pmb{e}_z\sin{k_Ly}+E_{yx}\pmb{e}_x\cos{k_Ly}$, where $E_{\mu\nu}(\mu,\nu=x,y,z)$ are field components propagating in direction $\mu$ with polarization $\nu$, the wave number $k_L$ of $\pmb{E}_x$ and $\pmb{E}_y$ is approximately same, and all other irrelevent phases in light fields are ignored. The frequency difference between the two sets of beams compensates the Zeeman splitting as, $\omega_{xy}-\omega_{yx}=\delta_F$, $\omega_{xz}-\omega_{yz}^{(1)}=\delta_F-2\delta_{B}$, and $\omega_{xz}-\omega_{yz}^{(2)}=\delta_F+2\delta_{B}$, in which $\delta_F$ is the initial hyperfine splitting between $\ket{F}$ and $\ket{F+1}$ (about several GHz).

The lattice potential $\hat{V}_{latt} \propto (\boldsymbol{E}_{x}^* \cdot \boldsymbol{E}_{x}+\boldsymbol{E}_y^* \cdot \boldsymbol{E}_y)$ is generally anisotropic in the x-y plane and forms a spin-dependent square lattice, taking the form $\hat{V}_{latt}(x,y)=(V_x\otimes\pmb{1}+\delta V_x\gamma_{3})\cos^2(k_Lx)+(V_y\otimes\pmb{1}+\delta V_y\gamma_{3})\cos^2(k_Ly)$ (see Appendix \ref{raman}). Without loss of generality, we consider the isotropic case and take $V_x=V_y=V_0$, and $\delta V_x=\delta V_y=\delta_V$ thereafter. Then our setting is described by the following Hamiltonian,
\begin{equation}
\begin{aligned}
H&=[\frac{\pmb{p}^2}{2m_{a}}\otimes\pmb{1}+\hat{V}_{latt}(\pmb{r})]+\mathcal{M}_{1}(\pmb{r})\gamma_{1}\\
&+\mathcal{M}_{2}(\pmb{r})\gamma_{2}+m_{z}\tau_3,\label{ham}
\end{aligned}
\end{equation}
where $\gamma^1=\sigma_1 \otimes \tau_0$, $\gamma^2=\sigma_2 \otimes \tau_2$, $\gamma^3=\sigma_3 \otimes \tau_0$ are three real Dirac matrices satisfying the Clifford algebra, with $\sigma_i$ and $\tau_i$ being two sets of the Pauli matrices, $\sigma_0$ and $\tau_0$ being the $2\times 2$ identity matrix, and $\pmb{1}=\sigma_0 \otimes \tau_0$. $m_a$ is the atomic mass and $m_z$ is brought by slightly tuning Zeeman field strength. The Raman potential $\mathcal{M}_{1}(\pmb{r})$ and $\mathcal{M}_{2}(\pmb{r})$ read as $\mathcal{M}_{1}(\pmb{r})=M_{10}\sin(k_Lx)\cos(k_Ly)$ with $M_{10}\propto(\frac{1}{\Delta_{1}}-\frac{1}{\Delta_{2}})E_{xy}E_{yx}$, and $\mathcal{M}_{2}(\pmb{r})=M_{20}\sin(k_Ly)\cos(k_Lx)$ with $M_{20}\propto(\frac{1}{\Delta_{1}}-\frac{1}{\Delta_{2}})E_{xz}E_{yz}$, here we set $E_{yz}\equiv E_{yz}^{(1)}=E_{yz}^{(2)}$.

We further take that fermions occupy the lowest s-orbitals $\phi_{s,\sigma\tau}$ ($\sigma=e,g$, $\tau=\uparrow, \downarrow$), and consider only the nearest-neighbor hoppings. Then we derive a tight-binding Hamiltonian,
\begin{equation}
\begin{aligned}
 H_s&=\sum_{\langle \mathbf{i}\mathbf{j}\rangle,\sigma\neq \sigma',\tau} t_1^{\mathbf{i}\mathbf{j}} c_{\mathbf{i},\sigma,\tau}^\dagger c_{\mathbf{j},\sigma',\tau}+ \sum_{\langle \mathbf{i}\mathbf{j}\rangle,\sigma\neq \sigma'} t_2^{\mathbf{i}\mathbf{j}} (c_{\mathbf{i},\sigma,\uparrow}^\dagger c_{\mathbf{j},\sigma',\downarrow}\\
&-c_{\mathbf{i},\sigma,\downarrow}^\dagger c_{\mathbf{j},\sigma',\uparrow})+\sum_{\langle \mathbf{i}\mathbf{j}\rangle,\sigma,\tau}t_{\sigma}^{\mathbf{i}\mathbf{j}} c_{\mathbf{i},\sigma,\tau}^\dagger c_{\mathbf{j},\sigma,\tau}\\
& +\sum_{\mathbf{i}}m_z(n_{\mathbf{i},\uparrow}-n_{\mathbf{i},\downarrow})+\delta_V(n_{\mathbf{i},e}-n_{\mathbf{i},g}),
\end{aligned}
\end{equation}
where $c_{\mathbf{i},\sigma,\tau}^\dagger$ ($c_{\mathbf{i},\sigma,\tau}$) is the creation (annihilation) operator with $\mathbf{i}=(i_x,i_y)$ denoting the lattice sites, the notation $\langle \mathbf{i}\mathbf{j}\rangle$ runs over all nearest-neighbor sites, and $n_{\mathbf{i},\sigma(\tau)}=\sum_{\tau(\sigma)}c_{\mathbf{i},\sigma,\tau}^\dagger c_{\mathbf{i},\sigma,\tau}$ is the particle number operator. The  strengths of related spin-flipped nearest-neighbour hoppings are given by:
\begin{equation}
\begin{aligned}
t_1^{\vec{i}\vec{j}}&=\int d^2\pmb{r} \phi^{(i)}_{s,e\uparrow}[M_{10}\sin{(k_Lx)}\cos{(k_Ly)}]\phi^{(j)}_{s,g\uparrow},\\
t_2^{\vec{i}\vec{j}}&=\int d^2\pmb{r} \phi^{(i)}_{s,e\uparrow}[M_{20}\sin{(k_Ly)}\cos{(k_Lx)}]\phi^{(j)}_{s,g\downarrow},
\end{aligned}
\end{equation}
while the strengths of spin-conserved hoppings are given by:
\begin{equation}
\begin{aligned}
t_{e}^{\vec{i}\vec{j}}&=\int d^2\pmb{r} \phi^{(i)}_{s,e\sigma}[\frac{\pmb{p}^2}{2m_a}+V_+(\cos^2k_Lx+\cos^2k_Ly)] \phi^{(j)}_{s,e\sigma},\\
t_{g}^{\vec{i}\vec{j}}&=\int d^2\pmb{r} \phi^{(i)}_{s,g\sigma}[\frac{\pmb{p}^2}{2m_a}+V_-(\cos^2k_Lx+\cos^2k_Ly)] \phi^{(j)}_{s,g\sigma},
\end{aligned}
\end{equation}
$V_{\pm}=V_0\pm \delta_V$. We further perform a gauge transformation $c_{\mathbf{i},g,\tau}\rightarrow e^{-i\pi (i_x+i_y)}c_{\mathbf{i},g,\tau}$ to absorb the staggered sign in the spin-flipped hopping terms.  After Fourier transformation, the Bloch Hamiltonian in $\pmb{k}$ space reads $H_{s,\pmb{k}}=\sum_{\pmb{k}} c_{\pmb{k}}^\dagger \mathcal{H}(\pmb{k}) c_{\pmb{k}}$, with $c_{\pmb{k}}=(c_{\pmb{k},e\uparrow},c_{\pmb{k},e\downarrow},c_{\pmb{k},g\uparrow},c_{\pmb{k},g\downarrow})^{\rm T}$, and $\mathcal{H}({\pmb{k}})=m_z \tau_{3}+\pmb{d}\cdot\pmb{\gamma}$. Here we have $d_1=2t_1\sin{k_y}$, $d_2=2t_2\sin{k_x}$, $d_3=\delta_V-2t_3(\cos{k_x}+\cos{k_y})$ and $d_0=2t_0(\cos{k_x}+\cos{k_y})$, where $t_3=(t_e+t_g)/2$ and $t_0=(t_e-t_g)/2$.

The Hamiltonian $\mathcal{H}({\pmb{k}})$ is $\mathcal{PT}$ invariant, $(\mathcal{PT})\mathcal{H}({\pmb{k}})(\mathcal{PT})^{-1}=\mathcal{H}({\pmb{k}})$, with $\mathcal{PT}=\mathcal{K}$ ($\mathcal{K}$ is the complex conjugate operator), which guarantees that the reality condition holds for each Bloch band. $\mathcal{H}({\pmb{k}})$ also respects the single $\mathcal{P}$ and $\mathcal{T}^2=1$ symmetry, with $\mathcal{P}=\gamma_3$ and $\mathcal{T}=\gamma_3\mathcal{K}$, and additional $\mathcal{T}^2=-1$ only when $m_z=0$. For more discussions on extra symmetries, see Appendix \ref{symmetry}. The four Bloch bands are $\varepsilon_{n,\pm}(\pmb{k})=d_0+(-1)^{(n)}\sqrt{(\sqrt{d_1^2+d_3^2}\pm m_z)^2+d_2^2}$ ($n=1,2$). The valence bands $\varepsilon_{\pmb{k},\pm}$ are globally degenerate for $m_z=0$, otherwise they only have two accidental degeneracy points residing at $\pmb{K}_{\pm}=(\pm\cos^{-1}[\delta_V/(2t_3)-1],0)$ or $(\pm\cos^{-1}[\delta_V/(2t_3)+1],\pi)$. The topology of the system is characterized by the second Stiefel-Whitney class \cite{YXZhao2017,Slager20201},\begin{equation}\nu=\frac{1}{4\pi}\int_{T^2} \mathrm {Tr}[I\mathcal{F}_R] ~\mathrm{d}k_x\mathrm{d}k_y\quad \mathrm{mod}~2,\end{equation}where $I=-i\sigma_2$ is the generator of the SO(2) group, and $\mathcal{F}_R=\nabla_{\pmb{k}}\times\pmb{A}(\pmb{k})$ is the non-Abelian Berry curvature for the real bundle with $\pmb{A}_{mn}(\pmb{k})=\braket{u^m_{\pmb{k}}|\nabla_{\pmb{k}}|u^n_{\pmb{k}}}$ called the real Berry connection, and $\ket{u^n_{\pmb{k}}}$ is a real occupied Bloch state. One can verify that we have a nontrivial $\nu$ for $\delta_V/t_3\in (-4,0)\cap(0,4)$, and $m_z$ does not affect the topological number as it can not close the bulk gap.

We numerically solve the continuous Hamiltonian (\ref{ham}) using a Fourier series expansion of a Bloch function (see Appendix \ref{full-band}). The structure of the lowest s-bands is shown in Fig. \ref{fig1}(c). The inversion of the spin polarization $\braket{\gamma_3}$ in the valence and conduction bands reveals the band repulsion induced by the SOC, indicating the topological nature of our model. This inspires a method to detect the topology of our model, as we will discuss later. Furthermore, we map out the phase diagram of the continuous Hamiltonian (\ref{ham}) based on the second SW class of the valence bands. The second SW class is numerically calculated using the four parities of the Bloch states at high symmetric momentum points. We find a large nontrivial regime on the phase plane in Fig. \ref{fig1}(d), which is accessible with current technology. Two regimes with $\nu=\pm1$ are topologically equivalent due to their $\mathbb{Z}_2$ nature, but they are separated by a gap-closing event where $V_0=\delta_V=0$.

\begin{figure}[htbp]
	\centering
	\includegraphics[width=\textwidth]{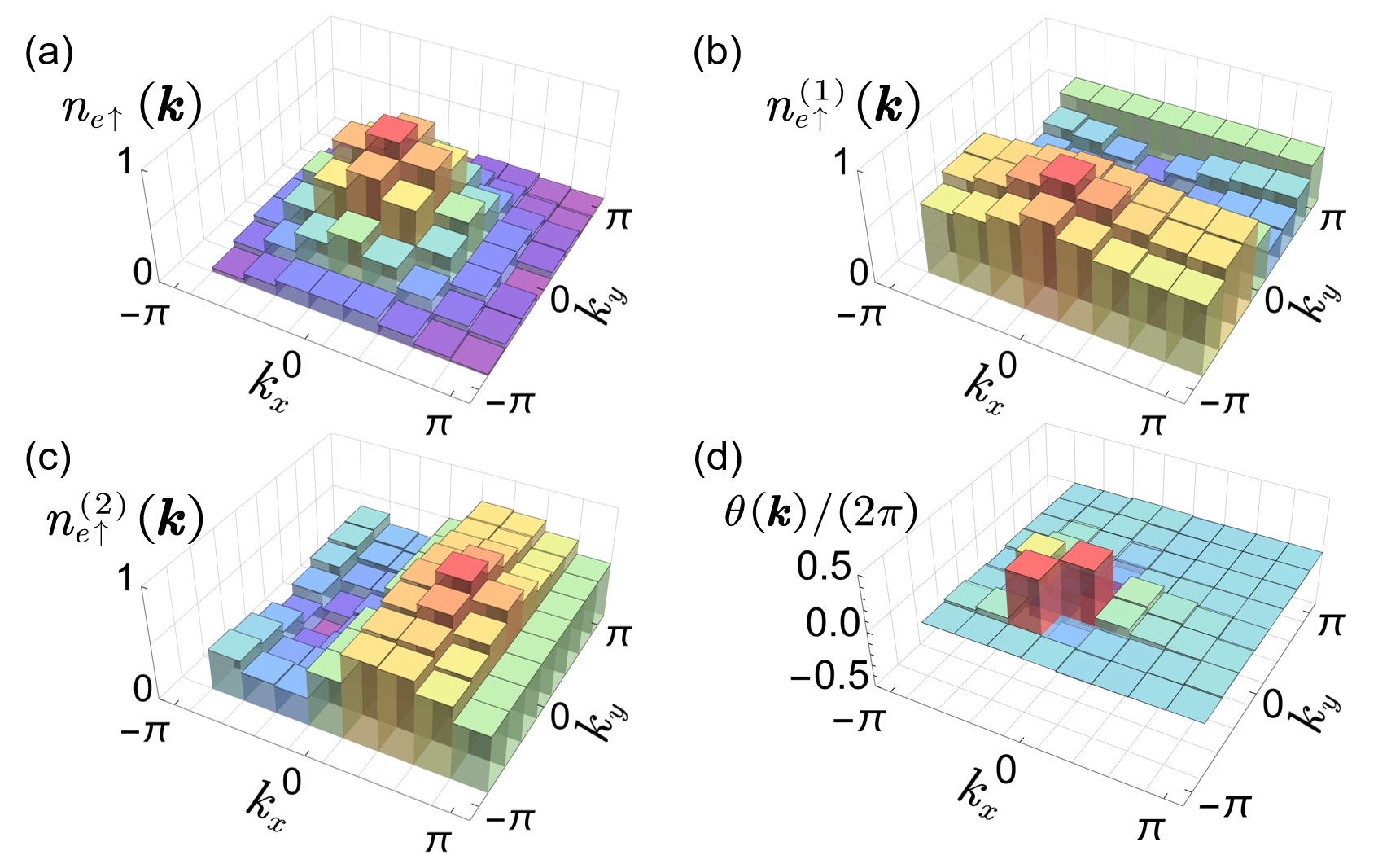}
  \caption{Density distributions of the spin states for $n_{e\uparrow}$ with (a) no pulses added, (b) added $T_1$, and (c) added $T_2$ in momentum space for a finite periodic lattice with $8 \times 8$ sites. The total density at each point is $\sum_{i} n_{i}(\pmb{k})=2$, corresponding to a half-filling case. (d) Extracted real Berry curvature at discretized Brillouin zone. The summation gives the second SW class $\nu=1$. Parameters are chosen as $t_2/t_1=t_3/t_1=1$, $\delta_V/t_1=2$, and $m_z=0$.}\label{fig2}
\end{figure}

\section{Equilibrium detection}\label{detect_eq}

We now proceed to the direct probe of the second SW class via an O(2) link method. We assume that the system is prepared in its ground state with a half-filling condition, 
\begin{equation}
\ket{G}=\prod_{\mathbf{k}} a^\dagger_{\mathbf{k},1+}\prod_{\mathbf{k}} a^\dagger_{\mathbf{k},1-}|0\rangle,
\end{equation}
where $a^\dagger_{\mathbf{k},1\pm}$ are occupied eigenmodes, related to Bloch states expressed in the spin basis by $\ket{u^{\alpha}_{\mathbf{k}}}=a^\dagger_{\mathbf{k},\alpha}\ket{0}=\sum_{\beta}[u^{\alpha}_{\mathbf{k}}]^{\beta} c^\dagger_{\mathbf{k},\beta}\ket{0}$, in which $[u^{\alpha}_{\mathbf{k}}]^\beta$ is the $\beta$ component with a real value.

With the system cooling down to the ground state, we then turn off the optical potential and perform TOF imaging. However, the measurement of quasi-momentum distribution only gives the amplitude information $n_{\beta}(\mathbf{k})=\sum_{\pm}|[u^{1\pm}_{\mathbf{k}}]^{\beta}|^2$ with the mixing contributions of the two degenerate bands. To separately extract the Bloch states, we apply an impulsive pulse right before TOF to induce a rotation between different spin components. For instance, a rotation $T_1=e^{i\frac{\pi}{4}\sigma_2}$ transforms $n_{e\uparrow}(\mathbf{k})$ to $n^{(1)}_{e\uparrow}(\mathbf{k})=[n_{e\uparrow}(\mathbf{k})+n_{g\uparrow}(\mathbf{k})]/2$, while another rotation $T_2=e^{i\frac{\pi}{4}\sigma_3\otimes \tau_2}$ transforms $n_{e\uparrow}(\mathbf{k})$ to $n^{(2)}_{e\uparrow}(\mathbf{k})=[n_{e\uparrow}(\mathbf{k})+n_{g\downarrow}(\mathbf{k})]/2$. In this way, we obtain the full tomography of the Bloch states in $\mathcal{H}(\mathbf{k})$ when $m_z=0$. Then we discretize the TOF image and define a connection matrix of Bloch states at the near quasi-momentum pixel, $[\theta^{x(y)}_{\mathbf{k}}]$ by $[\theta^{x(y)}_{\mathbf{k}}]^{\alpha\beta}=\braket{u^{\alpha}_{\mathbf{k}+\delta\mathbf{k_x(k_y)}}|u^{\beta}_{\mathbf{k}}} $, in the spirit of a real Wilson loop. An O(2) link is given by\begin{equation}\mathcal{W}_{\mathbf{k}}=[\theta^{y}_{\mathbf{k}}]^{-1}[\theta^{x}_{\mathbf{k}+\delta\mathbf{y}}]^{-1}[\theta^{y}_{\mathbf{k}+\delta\mathbf{x}}][\theta^{x}_{\mathbf{k}}]. \label{link}\end{equation}$[\theta^{x(y)}_{\mathbf{k}}]$ at each quasi-momentum $\mathbf{k}$ corresponds to an $\mathrm{SO}(2)$ Berry rotation with $\mathrm{det}[\theta]=+1$ or $-1$, due to the discontinuity of the $\mathrm{O}(2)$ group. Therefore, in the most general case, $\mathcal{W}_{\mathbf{k}}$ is gauge-covariant. Under a local gauge transformation $\mathcal{O}^{-1}_{\mathbf{k}}\mathcal{W}_{\mathbf{k}}\mathcal{O}(\mathbf{k})$ with $\mathrm{det}[\mathcal{O}]=-1$, the sign of $\mathcal{W}_{\mathbf{k}}$ changes. Therefore, we need to apply a parallel transport gauge to fix the orientation of the O(2) link \cite{Soluyanov2012}. After doing this, we build a gauge-independent field by $\mathcal{F}_{xy}=i\ln\mathcal{W}=i\theta^{tot}_{\mathbf{k}}\sigma_2$, corresponding to a discrete version of the non-Abelian real Berry curvature. The Euler class is then calculated by summing them up in the Brillouin zone: $\nu=\frac{1}{2\pi}\sum_{\mathbf{k}}\theta^{tot}_{\mathbf{k}}$.

To verify our method, we simulate the experimental signals by diagonalizing the real-space Hamiltonian on a finite lattice and using Fourier transformation \cite{LMDuan2006,DLDeng2014}, as shown in Fig. \ref{fig2}. To simulate realistic experiments, we add a global harmonic trap $V_{\text{trap}}=\frac{1}{2m_a}\omega^2 r^2$, which is parameterized by $\frac{\mu_T}{t_1}=\frac{1}{2m_a}\omega^2 a^2$. We also compare the results under different boundary conditions in Appendix \ref{table} and summarize them in Table \ref{table1}, which agree well with each other. Numerical calculations show that our method works well even for lattices with very small sites and near the phase boundary.

\begin{table}[ht]
\caption{Numerical results for topological index $\nu$ in different realistic conditions (with periodic or open boundaries, with or without trap). The size of the lattice is uniformly set to be $8\times 8$. The depth of harmonic trap is given by $\mu_T/t_1=0.01$. Other parameters are $t_2/t_1=t_3/t_1=1$, $m_z/t_1=0.4$.}
\centering
\begin{tabular}{c c c c c}
\hline\hline
$\delta_V/t_1$  &  Periodic  &  Open  &  Periodic+Trap  &  Open+Trap\\ [0.5ex]
\hline
2 & 1.000 & 1.000 & 1.000 & 0.999 \\ 
8 & 0 & 0 & 0 & 0\\
\hline 
\end{tabular}
\label{table1}
\end{table}

\begin{figure}[htbp]
	\centering
	\includegraphics[width=\textwidth]{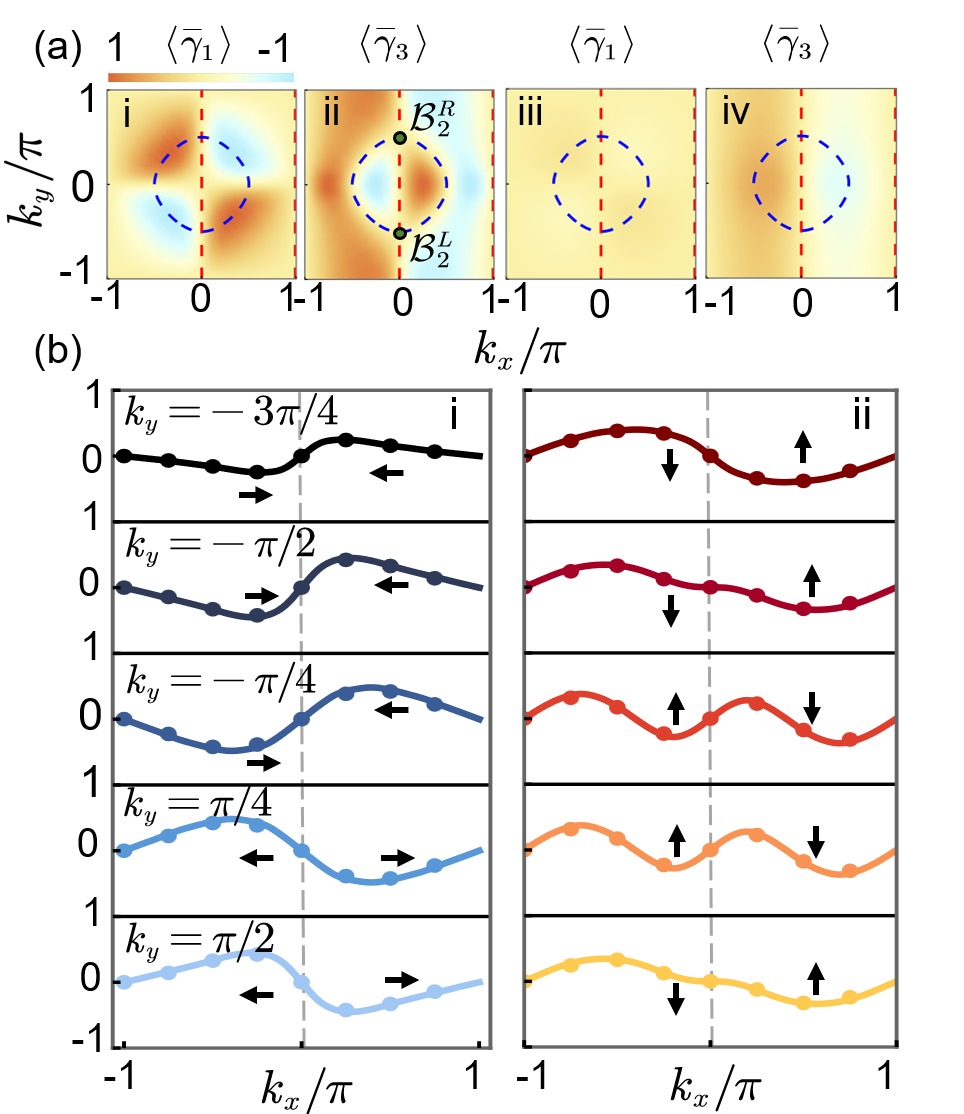}
  \caption{(a)Time-averaged spin polarization calculated using Bloch Hamiltonian, for \romannumeral1. $\braket{\overline{\gamma_1}}$ and \romannumeral2.  $\braket{\overline{\gamma_3}}$ in the topological phase ($\delta_V=1$); \romannumeral3.  $\braket{\overline{\gamma_1}}$ and \romannumeral4.  $\braket{\overline{\gamma_3}}$ in the trivial phase ($\delta_V=4$. Quench axis is $\gamma_0=\gamma_2$, with a ring-shaped BIS defiend by $d_3(\pmb{k})=0$, and a line-shaped BIS defined by $d_2(\pmb{k})=0$. (b)The sector of the TASP in topological phase calculated using Bloch Hamiltonian (lines) and finite real space lattice (solid circles). The arrows show the rotation directions of the spin vector $\pmb{\gamma}=(\braket{\overline{\gamma_1}},\braket{\overline{\gamma_3}})$ across the BIS (along $+k_x$ direction). For the finite system, the size is $8\times8$ with an extra harmonic trap with $\mu_T/t_1=0.01$. The evolution time after quenching is $T/t_1=20$. Here we take $\hbar=1$.}\label{fig3}
\end{figure}

\section{Dynamical detection}\label{detect_dyn}

The equilibrium detection requires the preparation of a nontrivial ground state. Next, we propose to probe the topology by quench dynamics (applied to the case where $m_z=0$), which starts from a trivial initial state and simply detects the outcome. The system is initialized in the deeply trivial regime with a large bias, then quenched to a target post-Hamiltonian. The initial state is thus fully polarized in the ground state of the flattened pre-quench Hamiltonian $H_0=m_i\gamma_i$ (the polarization axis can be arbitrary, but we specifically chose $i=2$ to illustrate our method). The time-averaged spin polarization (TASP) $\langle\overline{\gamma_i}\rangle$ at $\pmb{k}$ is given by:

\begin{equation}
\begin{aligned}
\braket{\overline{\gamma_i}}(\pmb{k})&=\frac{1}{T}\int_{0}^{T} \braket{\Psi^0|c^\dagger_{\pmb{k}}[e^{iH^t_{\pmb{k}}t}\gamma_ie^{-iH^t_{\pmb{k}}t}] c_{\pmb{k}}|\Psi^0}\\
&=-d_i(\pmb{k})d_2(\pmb{k})/[\varepsilon(\pmb{k})-d_0(\pmb{k})]^2,\label{longtimeave}
\end{aligned}
\end{equation}
where $H^t_{\mathbf{k}}=\sum_i d_i(\mathbf{k})\cdot\gamma_i$ is the post-quench Hamiltonian, and $T$ is the evolution time after the quench. We plot the numerical results in Fig. \ref{fig3}. The TASP $\mathbf{\gamma}\equiv (\langle\overline{\gamma_1}\rangle,\langle\overline{\gamma_3}\rangle)$ vanishes on a reduced structure called the BIS ($\mathcal{B}_1$), defined by $\mathcal{B}_1=\{\mathbf{k}|d_2(\mathbf{k})=0\}$. For the case with $\gamma_2$ as the quench axis, $\mathcal{B}_1$ is simply $k_x=0,\pi$. Furthermore, a certain component $\braket{\overline{\gamma_i}}$ of the TASP also vanishes on $\mathcal{B}'_1=\{\mathbf{k}|d_i(\mathbf{k})=0\}$, and its intersection with $\mathcal{B}_1$ gives rise to a higher-order BIS ($\mathcal{B}_2$) defined by $\mathcal{B}_2=\{\mathbf{k}|d_2=d_i=0\}$. As shown in Fig. \ref{fig3}(b), the spin vector $\mathbf{\gamma}$ exhibits nontrivial winding behavior across $\mathcal{B}_1$ in the topological case. This feature is captured by a field $\mathbf{g}(\mathbf{k})=\frac{1}{\mathcal{N}_{\mathbf{k}}}\partial_{\mathbf{k}_\perp} \mathbf{\gamma}$, where $\mathbf{k}_\perp$ denotes the momentum perpendicular to $\mathcal{B}_1$, and $1/\mathcal{N}_{\mathbf{k}}$ is the normalization factor. We show that the winding number of the field $\mathbf{g}(\mathbf{k})$ on the submanifold $\mathcal{B}_1$, $w_1=\sum_{i}\int_{\mathcal{B}^{(i)}_1}\mathrm{d}\mathbf{k}~ \mathbf{g}(\mathbf{k})\cdot\mathrm{d}\mathbf{g}(\mathbf{k})$, is equivalent to the second Stiefel-Whitney class (for a brief proof, see Appendix. \ref{bis}). We can understand this connection through the gauge transition between the two patches divided by the BIS \cite{YXZhao2016}. This correspondence builds a non-Abelian version of the dynamical bulk-surface duality for our system. The detection can even be further reduced to the second-order BIS ($\mathcal{B}_2$). The topology then relies on the parity of $\mathcal{B}_2$, $\nu=\frac{1}{2}\sum_{\mathcal{B}_2}\mathrm{sgn}(d_{3,L})-\mathrm{sgn}(d_{3,R})$.

\section{Conclusions}
In summary, we have proposed a feasible scheme to realize a four-band $\mathcal{PT}$-symmetric SWI in an optical Raman lattice, along with two different methods to detect nontrivial topology in our model. The proposed realization is discussed based on a natural Raman lattice approach, which is suitable for all alkali atoms with half-integer nuclear spins. Further detection methods are given for both the equilibrium and non-equilibrium cases. Through mathematical derivation and numerical simulation, we show the equivalence of some variations of the topological index in SWI and address the validity of these methods under realistic experimental imperfections, such as limited system size, different boundary conditions, and the existence of an extra harmonic potential. Furthermore, these detection methods are not limited by $\mathbb{Z}_2$ classification and can be directly applied to TEPs. Our proposed system would provide a promising platform for elucidating the exotic physics of SWI that is elusive in nature and may be realized with various artificial quantum systems~\cite{Goldman2014a,DWZhang2018,XTan2018_PRL,XTan2019_PRL,SLZhu2006_PRL,CMonroe2021,Georgescu2014,FMei2020_PRL}.

\acknowledgments

We thank  Y.Q. Zhu, Zhen Zheng, and S. L. Zhu for helpful discussions. This work was supported by the National Natural Science Foundation of China (Grant No. 12074180).

\begin{appendix}
\section{Symmetries in realized model}\label{symmetry}

In this section we give some discussions about extra symmetries in the proposed $(\mathcal{PT})^2=1$ symmetric model. We recall the form of the Bloch Hamiltonian,
\begin{equation}
\begin{aligned}
\mathcal{H}_{\pmb{k}}&=m_z\tau_3+2t_1\sin k_y \gamma_1+2t_2\sin k_x \gamma_2\\
&+[\delta_V-2t_3(\cos k_x+\cos k_y)] \gamma_3,\label{ham_k}
\end{aligned}
\end{equation}
with $2t_0(\cos k_x +\cos k_y)$ as a uniform hopping term. In all chosen sets of parameters, the Eq. (\ref{ham_k})  preserves $(\mathcal{PT})\mathcal{H}_{\pmb{k}}(\mathcal{PT})^{-1}=\mathcal{H}_{\pmb{k}}$ while preserving parity symmetry $\mathcal{P}\mathcal{H}_{\pmb{k}}\mathcal{P}^{-1}=\mathcal{H}_{-\pmb{k}}$ and time-reversal symmetry $\mathcal{T}\mathcal{H}_{\pmb{k}}\mathcal{T}^{-1}=\mathcal{H}_{-\pmb{k}}$ individually in which $\mathcal{P}=\gamma_3$ and $\mathcal{T}=\gamma_3\mathcal{K}$ symmetries with $\mathcal{P}^2=\mathcal{T}^2=1$. Spinless spatial symmetries other than $\mathcal{P}$ preserved when $m_z=0$ are mirror symmetries $M_{x,y}$ and $C_4$ symmetry, where $M_{x,y}^2=1, C_4^4=1$ and $C_4^2=\mathcal{P}$, given by:
\begin{equation}
\begin{aligned}
M_{x}\mathcal{H}_{(k_x,k_y)}M_{x}^{-1}&=\mathcal{H}_{(-k_x,k_y)},M_{x}=\sigma_0\otimes\tau_3,\\
M_{y}\mathcal{H}_{(k_x,k_y)}M_{y}^{-1}&=\mathcal{H}_{(k_x,-k_y)},M_{y}=\sigma_3\otimes\tau_3,\\
\end{aligned}
\end{equation}
and
\begin{equation}
C_4\mathcal{H}_{(k_x,k_y)}C_4^{-1}=\mathcal{H}_{(k_y,-k_x)},
\end{equation}
with
\begin{equation}
C_4=\begin{pmatrix}
\mathrm{i}\tau_2&0\\
0&-\mathrm{i}\tau_0
\end{pmatrix},
\end{equation}
where $\sigma_{i}$ and $\tau_{i}$ are Pauli matrices and $\sigma_0$ is the $2\times2$ identity matrix. When $m_z\neq0$ only the $C_4$ symmetry is broken.\par

\begin{figure}[htbp]
	\centering
	\includegraphics[width=\textwidth]{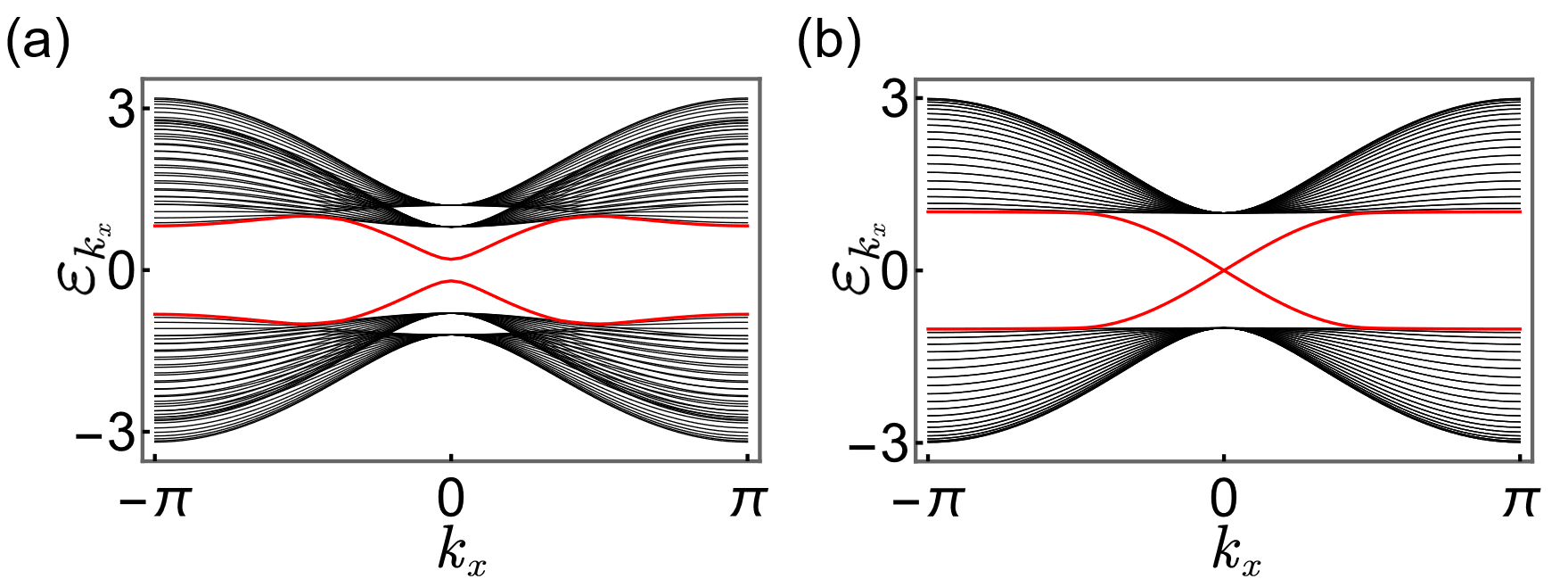}
  \caption{Energy band spectrum taken in periodic boundary along $x$ and open boundary along $y$ in topological nontrivial phase for $\delta_V/t_1=2$, $t_2/t_1=t_3/t_1=1$ and $t_0=0$ for (a) $m_z/t_1=0.4$ and (b)$m_z=0$. Edge states are labelled in red.}\label{fig_spec}
\end{figure}

When $m_z=t_0=0$, one can check that this Hamiltonian also preserves additional time-reversal symmetry $\mathcal{T}$ (TRS), paticle-hole symmetry $\mathcal{C}$ (PHS), and chiral symmetry $\mathcal{S}$ (CS), given by:
\begin{equation}
\begin{aligned}
\mathcal{T}&=\mathrm{i}\sigma_3\otimes\tau_2\mathcal{K}, \mathcal{T}^2=-1,\\
\mathcal{C}&=\mathrm{i}\sigma_1\otimes\tau_3\mathcal{K}, \mathcal{C}^2=+1,\\
\mathcal{S}&=\sigma_2\otimes\tau_1\quad, \mathcal{S}^2=+1.
\end{aligned}
\end{equation}
When $m_z\neq0$, PHS and CS and TRS with $\mathcal{T}^2=-1$ are all broken. In this case classification belongs to A\uppercase\expandafter{\romannumeral1} class in 2D with trivial topology. Thus nontrivial topology is considered to be brought by $\mathcal{PT}$ symmetry rather than $\mathcal{T}^2=+1$ symmetry alone. In FIG. \ref{fig_spec}, we show a comparison based on open boundary energy spectrum for $m_z\neq0$ and $m_z=0$, where the former one doesn't acquire a pair of zero-energy edge states at $k_x=0$.\par

\section{Details of Raman and optical potentials}\label{raman}

In this section we give details of light fields which generate the proposed potential and the Raman fields given in main text. The light fields are:
\begin{equation}
\begin{aligned}
\pmb{E}_x=E_{xz}\pmb{e}_z\cos{k_Lx}+\mathrm{i}E_{xy}\pmb{e}_y\sin{k_Lx},\\
\pmb{E}_y=\sum_{i=1,2}E_{yz}^{(i)}\pmb{e}_z\sin{k_Ly}+E_{yx}\pmb{e}_x\cos{k_Ly},
\end{aligned}
\end{equation}
where all $E_{\mu\nu}$ are real strength of light fields and $E^{(1)}_{yz}=E^{(2)}_{yz}=E_{yz}$. Setting quantization axis to be parallel to $\hat{z}$, light field $E_{xz,yz}$ drives $\pi$-transitions, and $E_{xy,yx}$ drives $\sigma_{\pm}$-transitions by decomposing into $\hat{e}_{\pm}$ basis:
\begin{equation}
\begin{aligned}
E_{xy}^{(-)}&=-\frac{i}{\sqrt{2}}E_{xy};~
E_{xy}^{(+)}=\frac{i}{\sqrt{2}}E_{xy};\\
E_{yx}^{(-)}&=\frac{1}{\sqrt{2}}E_{yx};~
E_{yx}^{(+)}=\frac{1}{\sqrt{2}}E_{yx}.
\end{aligned}
\end{equation}
To give a parameter estimation, we take further calculation in $^{133}\rm{Cs}$ atoms as an example. The chosen spinstates are $\ket{e_\uparrow}=\ket{F=4,mF=1}$, $\ket{e_\downarrow}=\ket{4,-1}$, $\ket{g_\uparrow}=\ket{3,-1}$, $\ket{g_\downarrow}=\ket{3,1}$ from ground state manifold $6S_{1/2}$. The hyperfine splitting is $\delta^S_F\approx 9.20\mathrm{GHz}$ in $6S_{1/2}$, $\delta^P_F\approx 1.17\mathrm{GHz}$ in $6P_{1/2}$ \cite{Cesium}. Extra magnetic field brings a energy shift estimated by $\delta_B\simeq 100\mathrm{MHz}$ which is much smaller than $\delta_F$, but still much larger than any parameters in effective Hamiltonian. As mentioned in main text, we have only D1 transition to be considered, which gives the Raman strength:
\begin{equation}
\begin{aligned}
M_{10}&=\frac{1}{16}\sqrt{\frac{5}{3}}\alpha_{D1}^2E_{xy}E_{yx}(\frac{1}{\Delta_1}-\frac{1}{\Delta_2}),\\
M_{20}&=\frac{1}{16}\sqrt{\frac{5}{3}}\alpha_{D1}^2E_{xz}E_{yz}(\frac{1}{\Delta_1}-\frac{1}{\Delta_2}),
\end{aligned}
\end{equation}
where $\Delta_2=\Delta_1+\delta^P_F$, $\alpha_{D1}\approx 3.19 ea_0$ is the related scalar polarizability with $a_0$ for Bohr radius.\par

Similar direct calculation gives spin-dependent potential of $\ket{e}$ and $\ket{g}$, given by:
\begin{equation}
\begin{aligned}
V_{ex}(x)&=\alpha_{D1}E_{xy}^2(\frac{13}{96}\frac{1}{\Delta_1-\Delta_F}+\frac{19}{96}\frac{1}{\Delta_2-\Delta_F})\sin^2(k_Lx)\\
&+\alpha_{D1}E_{xz}^2(\frac{5}{16}\frac{1}{\Delta_1-\Delta_F}+\frac{1}{48}\frac{1}{\Delta_2-\Delta_F})\cos^2(k_Lx),\\
V_{ey}(y)&=\alpha_{D1}E_{yx}^2(\frac{13}{96}\frac{1}{\Delta_1}+\frac{19}{96}\frac{1}{\Delta_2})\cos^2(k_Ly)\\
&+\alpha_{D1}E_{yz}^2(\frac{5}{8}\frac{1}{\Delta_1}+\frac{1}{24}\frac{1}{\Delta_2})\sin^2(k_Ly),\\
V_{gx}(x)&=\alpha_{D1}E_{xy}^2(\frac{11}{96}\frac{1}{\Delta_1}+\frac{7}{32}\frac{1}{\Delta_2})\sin^2(k_Lx)\\
&+\alpha_{D1}E_{xz}^2(\frac{1}{48}\frac{1}{\Delta_1}+\frac{5}{16}\frac{1}{\Delta_2})\cos^2(k_Lx),\\
V_{gy}(y)&=\alpha_{D1}E_{yx}^2(\frac{11}{96}\frac{1}{\Delta_1+\Delta_F}+\frac{7}{32}\frac{1}{\Delta_2+\Delta_F})\cos^2(k_Ly)\\
&+\alpha_{D1}E_{yz}^2(\frac{1}{24}\frac{1}{\Delta_1+\Delta_F}+\frac{5}{8}\frac{1}{\Delta_2+\Delta_F})\sin^2(k_Ly),\\\label{v_latt}
\end{aligned}
\end{equation}
where subscript indicates sublevel and direction. By discarding constants in Eq. (\ref{v_latt}) and forcing $\pmb{r}=(0,0)$ is at maximum, we reach a lattice potential $V_{latt}(x,y)=(V_x\pmb{1}+\delta V_x\gamma_{3})\cos^2(k_Lx)+(V_y\pmb{1}+\delta V_y\gamma_{3})\cos^2(k_Ly)$ which is further discussed in isotropic case in main text. Yet the  Eq. (\ref{v_latt}) holds a greater tunability of generated optical potential, such as a staggered spatial distribution of lattice sites for $\ket{e}$ and $\ket{g}$. \par

\section{Full-band Calculations}\label{full-band}

\begin{figure}[htbp]
	\centering
	\includegraphics[width=\textwidth]{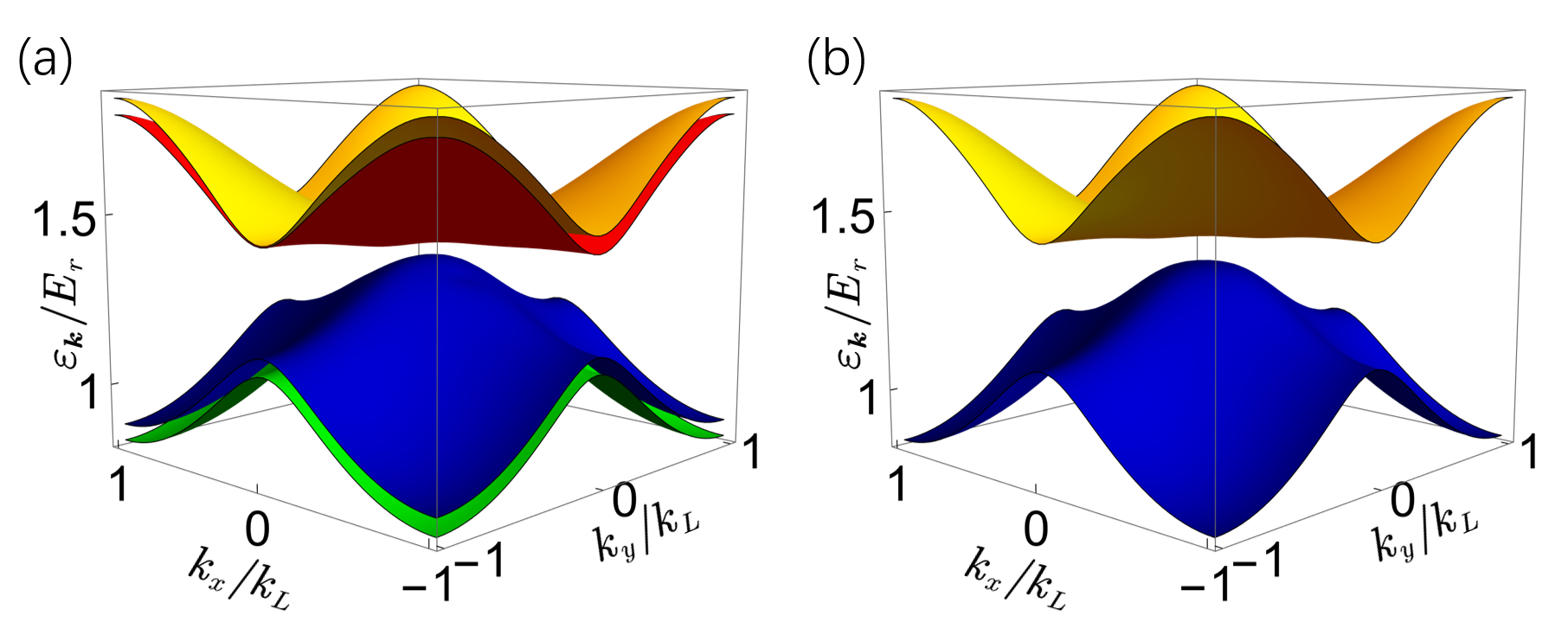}
  \caption{Band structures for four lowest bands taking part in effective TB model with (a) $m_z=0.05E_r$ and (b) $m_z=0$. Other parameters are $V_0=3E_r$, $\delta_V=0.3E_r$, $M_{10}=M_{20}=E_r$. The maximum order $N_{max}=5$.}\label{figapp}
\end{figure}

Here we introduce the derivation of results based on continuous Hamiltonian $\hatH$. We expand spatial periodic optical potential in the following way:
\begin{equation}
\begin{aligned}
V_e(x,y)&=V_{0e}+\frac{1}{4}V_{0e}(e^{\mathrm{i}2k_Lx}+e^{-\mathrm{i}2k_Lx}+e^{\mathrm{i}2k_Ly}+e^{-\mathrm{i}2k_Ly}),\\
V_g(x,y)&=V_{0g}+\frac{1}{4}V_{0g}(e^{\mathrm{i}2k_Lx}+e^{-\mathrm{i}2k_Lx}+e^{\mathrm{i}2k_Ly}+e^{-\mathrm{i}2k_Ly}),\\
\end{aligned}
\end{equation}
where we set $V_x=V_y=V_0$, $\delta V_x=\delta V_y=\delta_V$, and denotions $V_{0e}=V_0+\delta_V$ and $V_{0g}=V_0-\delta_V$. And Raman potential is also expressed by
\begin{equation}
\begin{aligned}
\mathcal{M}_1(x,y)&=\frac{1}{4\mathrm{i}}M_{10}(e^{\mathrm{i}k_Lx}-e^{-\mathrm{i}k_Lx})(e^{\mathrm{i}k_Ly}+e^{-\mathrm{i}k_Ly}),\\
\mathcal{M}_2(x,y)&=\frac{1}{4\mathrm{i}}M_{20}(e^{\mathrm{i}k_Lx}+e^{-\mathrm{i}k_Lx})(e^{\mathrm{i}k_Ly}-e^{-\mathrm{i}k_Ly}),
\end{aligned}
\end{equation}
in which the Raman potentials possess half of period of optical potential. In a limited area $S=L^2$ with side length $L$, an orthonormal set of wave function with quasimomentum $\pmb{k}=(k_x,k_y)$ is $\{\psi^{m,n}_{e}(\pmb{k})\ket{e_\uparrow},\psi^{m,n}_{e}(\pmb{k})\ket{e_\downarrow},\psi^{m,n}_{g}(\pmb{k})\ket{g_\uparrow},\psi^{m,n}_{g}(\pmb{k})\ket{g_\downarrow}\}$ with

\begin{equation}
\begin{aligned}
\psi^{m,n}_{e}(\pmb{k})&=\frac{1}{L}e^{\mathrm{i}(2mk_L+k_x)x}e^{\mathrm{i}(2nk_L+k_y)y},\\
\psi^{m,n}_{g}(\pmb{k})&=\frac{1}{L}e^{\mathrm{i}(2mk_L+k_L+k_x)x}e^{\mathrm{i}(2nk_L+k_L+k_y)y},
\end{aligned}
\end{equation}

where the phase difference between $\psi^{m,n}_{e}$ and $\psi^{m,n}_{g}$ is brought by Raman processes along $\pm\hat{e}_x\pm\hat{e}_y$ direction. Under a finite cut-off order $N_{max}$, such that $|m|,|n|\leq N_{max}$, Hamiltonian $\hat{H}$ has its matrix representation on this basis, and related eigen problem reads:
\begin{equation}
H\ket{\Psi^l_{\pmb{k}}}=\varepsilon^{l}_{\pmb{k}}\ket{\Psi^l_{\pmb{k}}},
\end{equation}
where eigenfunction of band index $l$ is then given by
\begin{equation}
\begin{aligned}
\ket{\Psi^l_{\pmb{k}}}=&\sum_{m,n,s=\uparrow,\downarrow}a^{l}_{m,n,s}(\pmb{k})\psi^{m,n}_{e}(\pmb{k})\ket{e_s}+\\
& b^{l}_{m,n,s}(\pmb{k})\psi^{m,n}_{g}(\pmb{k})\ket{g_s},
\end{aligned}
\end{equation}
with energy eigenvalue $\varepsilon^{l}_{\pmb{k}}$. Here we show complete s-band structure FIG. \ref{figapp}, with four lowest eigenvalues which is only partially given in main text. As for primarily giving the topological phase diagram from observables related to $\ket{\Psi^l_{\pmb{k}}}$, we consider the simple case where $m_z=0$. It's noticed that the Bloch Hamiltonian $\mathcal{H}_{\pmb{k}}$ commutes with $g_{02}$, which allows us to divide the Hamiltonian into two Chern sections:
\begin{equation}
\mathcal{H}_{\pmb{k}}=\mathcal{H}^+_{c}\otimes\sigma^+_2+\mathcal{H}^-_{c}\otimes\sigma^-_2,
\end{equation}
where $\sigma^\pm_2=\ket{\pm}\bra{\pm}$ with $\sigma_2\ket{\pm}=\pm\ket{\pm}$ is two projected subspace formed by eigenstates of $\sigma_2$.  Given originial $\mathcal{H}_{\pmb{k}}=\sum_{i=1,2,3}d_i\gamma_i$, the $\mathcal{H}^\pm_{c}=\sum_id^{\pm}_i\sigma_i$ with $\vec{d}_\pm=(d_1,\pm d_2,d_3)$, with preserved parity symmetry $\sigma_3\mathcal{H}^\pm_c(\pmb{k})\sigma_3=\mathcal{H}^\pm_c(-\pmb{k})$ and identical topological phase characterized by Chern number. Thus second SW class of $\mathcal{H}_{\pmb{k}}$ in this special case is mapped into Chern number in $\mathcal{H}_c$. We thus apply the minimal measurement for realized quantum anomalous Hall models, taking the criterion based on $\xi=\mathrm{sgn}(\left<\gamma_3\right>)$ at parity symmetric points $\Gamma=(0,0),X=(\pi,0),Y=(0,\pi),M=(\pi,\pi)$ \cite{XJLiu2013},
\begin{equation}
\nu=-\frac{\Theta}{2}\sum_i\xi_i,
\end{equation}
where $(-1)^{\Theta}=\prod_i\xi_i$. For the full-band calculation this can be settled from derived $\ket{\Psi^l_{\pmb{k}}}$:
\begin{equation}
\left<\gamma_3(\pmb{k})\right>_{l}=\left<\Psi^l_{\pmb{k}}|\gamma_3|\Psi^l_{\pmb{k}}\right>,
\end{equation}
which is spin-polarization along $\gamma_3$ for $l$ in four s-bands in zero temperature.

\section{Calculation of $\mathrm{O}_2$ links}\label{table}

In main text, we have shown the numerical results of extracting real Berry curvature under PBCs and $m_z=0$. For other conditions, we can still begin with a many-body ground state,
\begin{equation}
\ket{G}=\prod_{i=1}^{\mathcal{N}}a_{i}^{\dagger}\ket{0},
\end{equation}
where we directly denote creation operators of eigenmodes $\ket{\psi_i}\equiv a_{i}^{\dagger}\ket{0}$ in the order of eigenenergies $\varepsilon_i=\braket{\psi_i|H_s|\psi_i}$ from the lowest one, $\mathcal{N}=2\mathcal{L}^2$ is the number of particles maintaining half-filling with site number $\mathcal{L}$ along one side in square lattice. Each eigenmode can be formally expanded by a complete basis $\ket{\psi_i}\equiv a_{i}^{\dagger}\ket{0}=\sum_{\pmb{r},\alpha}[\psi_i]^{\pmb{r},\alpha}c^\dagger_{\pmb{r},\alpha}\ket{0}$ in which $[\psi_i]^{\pmb{r},\alpha}$ are coefficients related to site at $\pmb{r}$ and spin $\alpha=e_{\uparrow,\downarrow},g_{\uparrow,\downarrow}$. \par

To extract all discretized real Berry curvature, all the following distribution related to quasimomentum is essential for each condition:
\begin{equation}
n_{\alpha\beta}(\pmb{k})=\bra{G}c^\dagger_{\pmb{k},\alpha}c_{\pmb{k},\beta}\ket{G},
\end{equation}
which assumed to be achievable through tomography by extra pulses and TOF measurements in experiments. In our calculation, this is derived from Fourier transform from site representation, given by
\begin{equation}
\begin{aligned}
n_{\alpha\beta}(\pmb{k})&=\bra{G}c^\dagger_{\pmb{k},\alpha}c_{\pmb{k},\beta}\ket{G}\\
&=\bra{G}\frac{1}{\mathcal{L}}\sum_{\pmb{r}}e^{\mathrm{i}\pmb{k}\cdot\pmb{r}}c^\dagger_{\pmb{r},\alpha}\frac{1}{\mathcal{L}}\sum_{\pmb{r}'}e^{-\mathrm{i}\pmb{k}\cdot\pmb{r}'}c_{\pmb{r}',\beta}\ket{G}\\
&=\frac{1}{\mathcal{L}^2}\sum_{\pmb{r},\pmb{r}'}\bra{G}c^\dagger_{\pmb{r},\alpha}c_{\pmb{r}',\beta}\ket{G}e^{\mathrm{i}\pmb{k}\cdot(\pmb{r}-\pmb{r}')}\\
&=\frac{1}{\mathcal{L}^2}\sum_{\pmb{r},\pmb{r}',i}[\psi^*_i]^{\pmb{r},\alpha}[\psi_i]^{\pmb{r}',\beta}e^{\mathrm{i}\pmb{k}\cdot(\pmb{r}-\pmb{r}')},
\end{aligned}
\end{equation}
where $[\psi^*_i]^{\pmb{r},\alpha}$ denotes complex conjugate of this coefficient. In ideal case, $n(\pmb{k})$ is reduced to projector $P_{\pmb{k}}=\sum_{\pm}\ket{u^{1\pm}_{\pmb{k}}}\bra{u^{1\pm}_{\pmb{k}}}$ for occupied bands. \par

To fix the orientation that might be violated in this numerical process, we have two different ways. As in main text, we solve Bloch functions explicitly from northern gauge of stereographic representation, parameterized by
\begin{equation}
\begin{aligned}
\ket{u^{1+}_{\pmb{k}}}&=\frac{1}{\sqrt{x_{\pmb{k}}^2+y_{\pmb{k}}^2+1}}(x_{\pmb{k}},y_{\pmb{k}},-1,0),\\
\ket{u^{1-}_{\pmb{k}}}&=\frac{1}{\sqrt{x_{\pmb{k}}^2+y_{\pmb{k}}^2+1}}(-y_{\pmb{k}},x_{\pmb{k}},0,-1),
\end{aligned}
\end{equation}
in which two pulses are adequate to give full tomography of occupied Bloch states. For more general cases, i.e. conditions listed in Table. \ref{table1}, the parallel transport gauge is applied. This gauge is also called the cylinder gauge, in which wave functions are smooth inside the whole FBZ but the periodicity is only kept for one direction and broken for the other due to Wannier obstruction. We show some related result below. In contrast to the figure in main text which is derived from exhibiting accumulation of local real Berry curvature near the south pole, the distribution calculated by parallel transport gauge varies mildly.

\section{Bulk-Surface duality from deformation of $\mathcal{PT}$-symmetric Bloch Hamiltonian}\label{bis}

\begin{figure}[htbp]
	\centering
	\includegraphics[width=\textwidth]{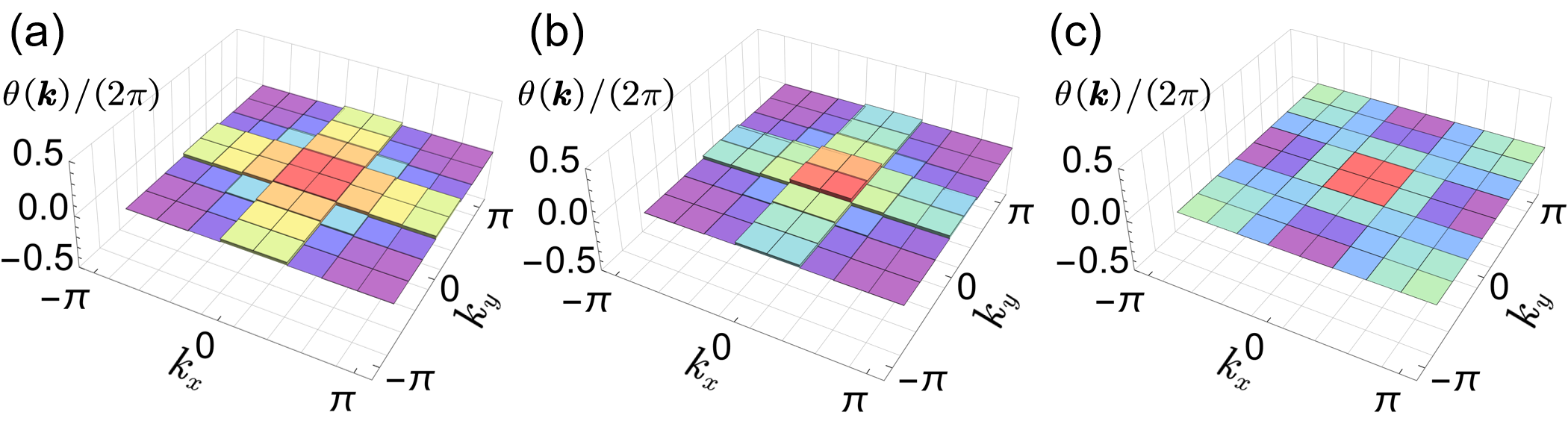}
  \caption{Discretized Berry curvature with parallel gauge. (a) $\nu=1$, with $t_2/t_1=t_3/t_1=1$, $\delta_V/t_1=2$ and $m_z=0$, under PBCs, which is same condition as in main text. (b)$\nu=0.999$, with $t_2/t_1=t_3/t_1=1$, $\delta_V/t_1=2$ and $m_z/t_1=0.4$, under OBCs. (c)$\nu=0$, with $t_2/t_1=t_3/t_1=1$, $\delta_V/t_1=8$ and $m_z/t_1=0.4$, under OBCs.}\label{fig_para}
\end{figure}

In this section, we give more discussions about extracting nontrivial topology from the BISs. To see this, for simplicity, we take a general spherical representation of the normalized Hamiltonian $H_{\pmb{k}}=\vec{d}\cdot\vec{\gamma}$ such that where $\vec{d}=(\sin\theta_{\pmb{k}},\cos\theta_{\pmb{k}}\sin\varphi_{\pmb{k}},\cos\theta_{\pmb{k}}\cos\varphi_{\pmb{k}})$ related with coordinate of a point on sphere. A gauge-invariant description of $\nu$ for $H_{\pmb{k}}=\vec{d}\cdot\vec{\gamma}$ can be given by:
\begin{equation}
\begin{aligned}
\nu&=\int_{T^2}-\frac{\mathrm{i}}{32\pi}\mathrm{Tr}[\tau_2H_{\pmb{k}}(\mathrm{d}H_{\pmb{k}})^2]\\
&=\frac{1}{8\pi}\int_{T^2}d^2\pmb{k}\epsilon_{ijk}d_{i}(\nabla_{\pmb{k}}d_j\times\nabla_{\pmb{k}}d_k),
\end{aligned}\label{nu_origin}
\end{equation}
where we use the relation $\mathrm{Tr}[\tau_2\gamma_i\gamma_j\gamma_k]=4\mathrm{i}\epsilon_{ijk}$. The topology for $H_{\pmb{k}}$ should be equivalent with another deformed Hamiltonian of $h(\pmb{k})=\vec{d'}\cdot\vec{\gamma}$ where $\vec{d'}=(\sin\Theta_{\pmb{k}},\cos\Theta_{\pmb{k}}\sin\varphi_{\pmb{k}},\cos\Theta_{\pmb{k}}\cos\varphi_{\pmb{k}})$ where $\Theta_{\pmb{k}}$ could be any monotomic function since both energy gap and $\mathcal{PT}$-symmetry are preserved \cite{LZhang2018}. Substituting into Eq. (\ref{nu_origin}), the topological invariant is written by $\nu=1/(4\pi)\oint d^2\pmb{k}\cos\Theta_{\pmb{k}}(\nabla_{\pmb{k}}\Theta_{\pmb{k}})\times(\nabla_{\pmb{k}}\varphi_{\pmb{k}})$. Under an extreme choice of $\Theta_{\pmb{k}}$ such that
\begin{equation}
\cos\Theta_{\pmb{k}}=\left\{
\begin{aligned}
1,\quad& \pmb{k}\in \mathrm{BISs}\\
0,\quad& otherwise
\end{aligned}\right.
\end{equation}
and $\nabla\Theta_{\pmb{k}}=\delta(\pmb{k}-\pmb{k}_{\mathrm{BISs}})$, the Eq. (\ref{nu_origin}) is then reduced into a line integral on closed 1D 1-BISs with exactly a winding number form $\nu=1/(2\pi)\oint_{\mathrm{BIS}}d\pmb{k}\nabla_{\pmb{k}}\varphi_{\pmb{k}}$ defined on the 1-BISs for $\gamma_1$. Since we don't acquire any explict form of Hamiltonian in this derivation, the choice of the reduced component $\gamma_1$ is indeed arbitrary. This can be iteratively done to reach higher-order BISs, as long as the deformation is valid.
\end{appendix}


\begin{thebibliography}{99}


\bibitem{DWZhang2018}
D.-W. Zhang, Y.-Q. Zhu, Y. X. Zhao, H. Yan and S.-L. Zhu,
Topological quantum matter with cold atoms,
Adv. Phys \textbf{67}, 253 (2018).

\bibitem{Goldman2014a}
N. Goldman, G. Juzeliūnas, P. Öhberg and I. B. Spielman,
Light-induced gauge fields for ultracold atoms,
Rep. Prog. Phys. \textbf{77}, 126401 (2014).

\bibitem{FMei2012_PRA} 
Feng Mei et al., 
Simulating $Z_2$ topological insulators with cold atoms in a one-dimensional optical lattice,
Phys. Rev. A \textbf{85}, 013638 (2012).

\bibitem{DWZhang2016_PRA} 
Dan-Wei Zhang et al., 
Quantum simulation of exotic PT-invariant topological nodal loop bands with ultracold atoms in an optical lattice, 
Phys. Rev. A \textbf{93}, 043617 (2016).

\bibitem{Hasan2010}
M. Z. Hasan and C. L. Kane,
Colloquium: topological insulators,
Rev. Mod. Phys. \textbf{82}, 3045 (2010).

\bibitem{XLQi2011}
X.-L. Qi and S.-C. Zhang,
Topological insulators and superconductors.
Rev. Mod. Phys. \textbf{83}, 1057 (2011).

\bibitem{CKChiu2016}
C.-K. Chiu, J. C. Y. Teo, A. P. Schnyder, and S. Ryu,
Classification of topological quantum matter with symmetries,
Rev. Mod. Phys. \textbf{88}, 035005 (2016).

\bibitem{XGWen2017}
X.-G. Wen,
Colloquium: Zoo of quantum-topological phases of matter,
Rev. Mod. Phys. \textbf{89}, 041004 (2017).

\bibitem{Schnyder2008}
A. P. Schnyder, S. Ryu, A. Furusaki and A. W. W. Ludwig,
Classification of topological insulators and superconductors in three spatial dimensions,
Phys. Rev. B. \textbf{78}, 195125 (2008).

\bibitem{Kitaev2009}
A. Kitaev,
Periodic table for topological insulators and superconductors,
AIP Conf. Proc. \textbf{1134}, 22-30 (2009).

\bibitem{Ryu2010}
S. Ryu, A. P. Schnyder, A. Furusaki and A. W. W. Ludwig,
Topological insulators and superconductors: tenfold way and dimensional hierarchy,
N. J. Phys. \textbf{12}, 065010 (2010).

\bibitem{Po2017}
H. C. Po, A. Vishwanath and H. Watanabe,
Symmetry-based indicators of band topology in the 230 space groups,
Nat. Commun. \textbf{8}, 50 (2017).

\bibitem{Kruthoff2017}
J. Kruthoff, J. de Boer, J. van Wezel, C. L. Kane and  R.-J. Slager,
Topological classification of crystalline insulators through band structure combinatorics,
Phys. Rev. X \textbf{7}, 041069 (2017).

\bibitem{Po2018}
H. C. Po, H. Watanabe and A. Vishwanath,
Fragile topology and wannier obstructions,
Phys. Rev. Lett. \textbf{121}, 126402 (2018).

\bibitem{Bradlyn2019}
B. Bradlyn, Z. Wang, J. Cano and B. A. Bernevig,
Disconnected elementary band representations, fragile topology, and Wilson loops as topological
indices: An example on the triangular lattice,
Phys. Rev. B. \textbf{99}, 045140 (2019).

\bibitem{Kooi2019}
S. H. Kooi, G. van Miert and C. Ortix,
Classification of crystalline insulators without symmetry indicators: atomic and fragile topological phases in twofold rotation symmetric systems,
Phys. Rev. B. \textbf{100}, 115160 (2019).

\bibitem{Bouhon2019}
A. Bouhon, A. M. Black-Schaffer and R.-J. Slager,
Wilson loop approach to fragile topology of split elementary band representations and topological
crystalline insulators with time-reversal symmetry,
Phys. Rev. B. \textbf{100}, 195135 (2019).

\bibitem{Slager20202}
A. Bouhon, T. Bzdusek, and R.-J. Slager,
Geometric approach to fragile topology beyond symmetry indicators,
Phys. Rev. B \textbf{102}, 115135 (2020).

\bibitem{Benalcazar2017}
W. A. Benalcazar, B. A. Bernevig, and T. L. Hughes,
Quantized electric multipole insulators,
Science, \textbf{357}, 61 (2017).

\bibitem{Benalcazar20172}
W. A. Benalcazar, B. A. Bernevig, and T. L. Hughes,
Electric multipole moments, topological multipole moment pumping, and chiral hinge
states in crystalline insulators,
Phys. Rev. B \textbf{96}, 245115 (2017).

\bibitem{Schindler2018}
F. Schindler, A. M. Cook, M. G. Vergniory, Z. Wang, S. S. P. Parkin, B. A. Bernevig  and T. Neupert,
Higher-order topological insulators,
Science Advances \textbf{4}, eaat0346 (2018).

\bibitem{BXie2021}
B. Xie, H.-X. Wang, X. Zhang, P. Zhan, J.-H. Jiang, M. Lu and Y. Chen,
Higher-order band topology,
Nat. Rev. Phys. \textbf{3}, 520-532 (2021).

\bibitem{CFang2015}C. Fang, Y. Chen, H.-Y. Kee, and L. Fu, Topological nodal line
semimetals with and without spin-orbital coupling, 
Phys. Rev. B \textbf{92}, 081201(R) (2015).


\bibitem{YXZhao2017}
Y. X. Zhao and Y. Lu,
PT-Symmetric Real Dirac Fermions and Semimetals,
Phys. Rev. Lett. \textbf{118}, 056401 (2017).

\bibitem{YXZhao2020}
K. Wang, J. X. Dai, L. B. Shao, S. A. Yang and Y. X. Zhao,
Boundary Criticality of $\mathcal{PT}$-Invariant Topology and Second-Order Nodal-Line Semimetals,
Phys. Rev. Lett. \textbf{125}, 126403 (2020).

\bibitem{Slager20201}
A. Bouhon, QuanSheng Wu, R.-J. Slager, H. Weng, O. V. Yazyev and T. Bzdusek,
non-Abelian reciprocal braiding of Weyl points and its manifestation in ZrTe,
Nat. Phys. \textbf{16}, 1137-1143 (2020).

\bibitem{Slager20203}
F. Nur Unal, A. Bouhon, and R.-J. Slager,
Topological euler class as a dynamical observable in optical lattices,
Phys. Rev. Lett. \textbf{125}, 053601 (2020).

\bibitem{JAhn2018}
J. Ahn, D. Kim, Y. Kim, and B.-J. Yang,
Band Topology and Linking Structure of Nodal Line Semimetals with Z2 Monopole Charges,
Phys. Rev. Lett. \textbf{121}, 106403 (2018).

\bibitem{JAhn2019}
J. Ahn, S. Park, and B.-J. Yang,
Failure of Nielsen-Ninomiya Theorem and Fragile Topology in Two-Dimensional Systems with Space-Time Inversion Symmetry: Application to Twisted Bilayer Graphene at Magic Angle,
Phys. Rev. X  \textbf{9}, 021013 (2019).

\bibitem{QWu2019}
Q. Wu, A. A. Soluyanov, and T. Bzdusek,
non-Abelian band topology in noninteracting metals,
Science \textbf{365}, 1273-1277 (2019).

\bibitem{Ezawa2021}
M. Ezawa,
Topological Euler insulators and their electric circuit realization,
Phys. Rev. B \textbf{103}, 205303(2021).

\bibitem{Takahashi2023}
R. Takahashi and T. Ozawa,
Bulk-edge correspondence of Stiefel-Whitney and Euler insulators through the entanglement spectrum and cutting procedure,
Phys. Rev. B \textbf{108}, 075129(2023).

\bibitem{YXZhao2016}
Y. X. Zhao, Andreas P. Schnyder, and Z. D. Wang,
Unified Theory of PT and CP Invariant Topological Metals and Nodal Superconductors,
Phys. Rev. Lett. \textbf{116}, 156402(2016).

\bibitem{Lim2023}
H. Lim, S. Kim, and B.-J. Yang,
Real Hopf insulator,
arXiv:2303.13078 (2023).

\bibitem{Bouhon2023}
A. Bouhon, Y.-Q. Zhu, R.-J. Slager, and G. Palumbo,
Second Euler number in four dimensional synthetic matter, 	
arXiv:2301.08827 (2023).



\bibitem{WZhao2022}
W. Zhao, Y.-B. Yang, Y. Jiang et al.,
Quantum simulation for topological Euler insulators,
Commun. Phys. \textbf{5}, 223 (2022).




\bibitem{SLZhu2007}
S.-L. Zhu, B. Wang, and L.-M. Duan,
Simulation and detection of Dirac fermions with cold atoms in an optical lattice
Phys. Rev. Lett. \textbf{98}, 260402 (2007).

\bibitem{LBShao2008}
L. B. Shao et al., Realizing and detecting the quantum Hall effect without Landau levels by using ultracold atoms,
Phys. Rev. Lett. \textbf{101}, 246810 (2008).

\bibitem{Tarruell2012}
L. Tarruell, D. Greif, T. Uehlinger, G. Jotzu and T. Esslinger,
Creating, moving and merging Dirac points with a Fermi gas in a tunable honeycomb lattice,
Nature \textbf{483}, 302-305 (2012).

\bibitem{DWZhang2020_SC} 
D. W. Zhang et al., 
Non-Hermitian topological Anderson insulators, 
Sci. China-Phys. Mech. Astron. \textbf{63}, 267062 (2020).

\bibitem{SLZhu2013_PRL}  S.-L. Zhu, Z.-D. Wang, Y.-H. Chan, and L.-M. Duan,  
Topological Bose-Mott Insulators in a One-Dimensional Optical Superlattice,
Phys. Rev. Lett. \textbf{110}, 075303 (2013).


\bibitem{DWZhang2020_PRB} 
Dan-Wei Zhang et al.,  
Skin superfluid, topological Mott insulators, and asymmetric dynamics in an interacting non-Hermitian Aubry-André-Harper model, 
Phys. Rev. B \textbf{101}, 235150 (2020).


\bibitem{Jotzu2014}
G. Jotzu, M. Messer, R. Desbuquois, M. Lebrat, T. Uehlinger, D. Greif and T. Esslinger,
Experimental realization of the topological Haldane model with ultracold fermions.
Nature \text{515}, 237-240 (2014).

\bibitem{BSong2019}
B. Song, C. He, S. Niu, L. Zhang, Z. Ren, X.-J. Liu and G.-B. Jo,
Observation of nodal-line semimetal with ultracold fermions in an optical lattice,
Nat. Phys. \textbf{15}, 911-916 (2019).

\bibitem{Minguzzi2022}
J. Minguzzi, Z. Zhu, K. Sandholzer, A.-S. Walter, K. Viebahn, and T. Esslinger,
Topological Pumping in a Floquet-Bloch Band
Phys. Rev. Lett. \textbf{129}, 053201 (2022).


\bibitem{Dalibard2011}
J. Dalibard, F. Gerbier, G. Juzeliūnas, P. Öhberg,
Colloquium: Artificial gauge potentials for neutral atoms,
Rev. Mod. Phys. \textbf{83}, 1523 (2011).




\bibitem{SLZhu2006}
S.-L. Zhu, H. Fu, C.-J. Wu, S.-C. Zhang, and L.-M. Duan,
Spin Hall Effects for Cold Atoms in a Light-Induced Gauge Potential,
Phys. Rev. Lett. \textbf{97}, 240401 (2006).

\bibitem{Beeler2013}
M. C. Beeler et al.,
The spin Hall effect in a quantum gas,
Nature \textbf{498}, 201 (2013).

\bibitem{YJLin2009}
Y.-J. Lin et al.,
Bose-Einstein Condensate in a Uniform Light-Induced Vector Potential,
Phys. Rev. Lett. \textbf{102}, 130401 (2009).

\bibitem{Ruseckas2005}
J. Ruseckas, G. Juzeliūnas, P. Öhberg, and M. Fleischhauer,
non-Abelian gauge potentials for ultracold atoms with degenerate dark states,
Phys. Rev. Lett. \textbf{95}, 010404 (2005).

\bibitem{Jaksch2003}
D. Jaksch and P. Zoller,
Creation of effective magnetic fields in optical lattices: the Hofstadter butterfly for cold neutral atoms,
New J. Phys. \textbf{5}, 56 (2003).

\bibitem{Goldman2009}
N. Goldman et al.,
non-Abelian optical lattices: Anomalous quantum Hall effect and Dirac fermions,
Phys. Rev. Lett. \textbf{103}, 035301 (2009).

\bibitem{Gorg2019}
F. Görg, K. Sandholzer, J. Minguzzi, R. Desbuquois, M. Messer and Tilman Esslinger,
Realization of density-dependent Peierls phases to engineer quantized gauge fields coupled to ultracold matter,
Nat. Phys. \textbf{15}, 1161-1167 (2019).

\bibitem{YJLin2011}
Y.-J. Lin, K. Jiménez-García and I. Spielman,
Spin–orbit-coupled Bose–Einstein condensates, 
Nature \textbf{471}, 83–86 (2011). 

\bibitem{Galitski2013}
V. Galitski, I. Spielman,
Spin–orbit coupling in quantum gases,
Nature \textbf{494}, 49–54 (2013).

\bibitem{ZCXu2022} 
Z. C. Xu et al., 
Gain/loss effects on spin-orbit coupled ultracold atoms in two-dimensional optical lattices, 
Sci. China Phys. Mech. Astron. \textbf{65}, 283011 (2022).

\bibitem{XFZhou2013}
X.-F. Zhou, G.-C. Guo, W. Zhang, and W. Yi,
Exotic pairing states in a Fermi gas with three-dimensional spin-orbit coupling,
Phys. Rev. A \textbf{87}, 063606 (2013).

\bibitem{XCui2014}
X. Cui and W. Yi,
Universal Borromean Binding in Spin-Orbit-Coupled Ultracold Fermi Gases,
Phys. Rev. X \textbf{4}, 031026 (2014).


\bibitem{BZWang2018}
B.-Z. Wang, Y.-H. Lu, W. Sun, S. Chen, Y. Deng and X.-J. Liu,
Dirac-, Rashba-, and Weyl-type spin-orbit couplings: Toward experimental realization in ultracold atoms,
Phys. Rev. A \textbf{97}, 011605(R) (2018).

\bibitem{JHZhang2022}
J.-H. Zhang, B.-B. Wang, F. Mei, J. Ma, L. Xiao, and S. Jia,
Topological optical Raman superlattices,
Phys. Rev. A \textbf{105}, 033310 (2022).

\bibitem{GLiu2010_PRA} 
G. Liu et al., 
Simulating and detecting the quantum spin Hall effect in the kagome optical lattice, 
Phys. Rev. A \textbf{82}, 053605 (2010).

\bibitem{DWZhang2012_PRA} 
Dan-Wei Zhang et al.,   
Macroscopic Klein tunneling in spin-orbit-coupled Bose-Einstein condensates, 
Phys. Rev. A \textbf{85}, 013628 (2012).

\bibitem{SLZhu2011_PRL} 
S. L. Zhu, L. B. Shao, Z. D. Wang, and L.M.Duan, 
Probing non-Abelian statistics of Majorana fermions in ultracold atomic superfluid,
Phys. Rev. Lett. \textbf{106}, 100404 (2011).

\bibitem{LHuang2016}
L. Huang et al., 
Experimental realization of two-dimensional synthetic spin–orbit coupling in ultracold Fermi gases,
Nat. Phys. \textbf{12}, 540 (2016).

\bibitem{JZLi2022_PRL} 
Jia-Zhen Li et al.,
Synthetic Topological Vacua of Yang-Mills Fields in Bose-Einstein Condensates, 
Phys. Rev. Lett. \textbf{129}, 220402 (2022).

\bibitem{ZWu2016}
Z. Wu et al.,
Realization of two-dimensional spin-orbit coupling for Bose-Einstein condensates,
Science \textbf{354}, 83 (2016).

\bibitem{ZYWang2021}
Z.-Y. Wang et al.,
Realization of an ideal Weyl semimetal band in a quantum gas with 3D spin-orbit coupling,
Science \textbf{372}, 271 (2021).

\bibitem{QXLv2023_PRA}  
Q.-X. Lv et al.,
Measurement of non-Abelian gauge fields using multiloop amplification, 
Phys. Rev. A \textbf{108}, 023316 (2023).


\bibitem{QXLv2021_PRL}   
Q.-X. Lv et al., 
Measurement of spin Chern numbers in quantum simulated topological insulators,
Phys. Rev. Lett. \textbf{127}, 136802 (2021).


\bibitem{Abanin2013}
D. A. Abanin, T. Kitagawa, I. Bloch, and Eugene Demler,
Interferometric approach to measuring band topology in 2D optical lattices,
Phys. Rev. Lett. \textbf{110}, 165304 (2013).

\bibitem{Atala2013}
M. Atala, M. Aidelsburger, J. T. Barreiro, D. Abanin, T. Kitagawa, E. Demler and I.Bloch,
Direct measurement of the Zak phase in topological Bloch bands,
Nat. Phys. \textbf{9}, 795-800 (2013).

\bibitem{Grusdt2014}
F. Grusdt, D. Abanin, and E. Demler,
Measuring $\mathbb{Z}_2$ topological invariants in optical lattices using interferometry,
Phys. Rev. A \textbf{89}, 043621 (2014).

\bibitem{Aidelsburger2015}
M. Aidelsburger, M. Lohse, C. Schweizer, M. Atala, J. T. Barreiro, S. Nascimbéne, N. R. Cooper, I. Bloch and N. Goldman,
Measuring the Chern number of Hofstadter bands with ultracold bosonic atoms, 
Nat. Phys. \textbf{11}, 162–166 (2015).

\bibitem{Dauphin2013}
A. Dauphin and N. Goldman,
Extracting the Chern number from the dynamics of a Fermi gas: implementing a quantum Hall bar for cold atoms,
Phys. Rev. Lett. \textbf{111}, 135302 (2013).

\bibitem{Alba2011}
E. Alba, X. Fernandez-Gonzalvo, J. Mur-Petit, J. K. Pachos, and J. J. Garcia-Ripoll,
Seeing Topological Order in Time-of-Flight Measurements,
Phys. Rev. Lett. \textbf{107}, 235301 (2011).

\bibitem{Hauke2014}
P. Hauke, M. Lewenstein, and André Eckardt,
Tomography of Band Insulators from Quench Dynamics,
Phys. Rev. Lett. \textbf{113}, 045303 (2014).

\bibitem{Li2016}
T. Li, L. Duca, M. Reitter, F. Grusdt, E. Demler, M. Endres, M. Schleier-Smith, I. Bloch, U.Schneider,
Bloch state tomography using Wilson lines,
Science \textbf{352}, 1094-1097 (2016).

\bibitem{RBLiu2015} 
Rui-Bin Liu et al., 
Directly probing the Chern number of the Haldane model in optical lattices,
Journal of the Optical Society of America B \textbf{32}, 2500 (2015). 

\bibitem{LZhang2018}
L. Zhang, L. Zhang, S. Niu and X.-J. Liu,
Dynamical classification of topological quantum phases,
Sci. Bull. \textbf{63}, 1385-1391 (2018).

\bibitem{XLYu2021}
X.-L. Yu, W. Ji, L. Zhang, Y. Wang, J. Wu, and X.-J. Liu,
Quantum Dynamical Characterization and Simulation of Topological Phases With High-Order Band Inversion Surfaces,
Phys. Rev. X Quantum \textbf{2}, 020320 (2021).


\bibitem{Soluyanov2012}
A. A. Soluyanov and D. Vanderbilt,
Smooth gauge for topological insulators,
Phys. Rev. B \textbf{85}, 115415 (2012).

\bibitem{LMDuan2006}
L. M. Duan,
Detecting Correlation Functions of Ultracold Atoms through Fourier Sampling of Time-of-Flight Images,
Phys. Rev. Lett. \textbf{96}, 103201 (2006).


\bibitem{DLDeng2014}
D.-L. Deng, S.-T. Wang, and L.-M. Duan,
Direct probe of topological order for cold atoms,
Phys. Rev. A \textbf{90}, 041601(R) (2014).


\bibitem{CMonroe2021} 
C. Monroe et al.,
Programmable quantum simulations of spin systems with trapped ions, 
Rev. Mod. Phys. \textbf{93}, 025001 (2021).

\bibitem{SLZhu2006_PRL} 
S. L. Zhu, C. Monroe, and L.M. Duan, 
Trapped Ion Quantum Computation with Transverse Phonon Modes, 
Phys. Rev. Lett. \textbf{97}, 050505 (2006).


\bibitem{FMei2020_PRL} 
F. Mei et al., 
Digital Simulation of Topological Matter on Programmable Quantum Processors,
Phys. Rev. Lett. \textbf{125}, 160503 (2020).

\bibitem{Georgescu2014} 
I. M. Georgescu, S. Ashhab, and F. Nori, 
Quantum simulation, 
Rev. Mod. Phys. \textbf{86}, 153 (2014).


\bibitem{XTan2018_PRL} 
X. Tan et al.,
Topological Maxwell Metal Bands in a Superconducting Qutrit, 
Phys. Rev. Lett. \textbf{120}, 130503 (2018).

\bibitem{XTan2019_PRL}
X. Tan et al.,
Experimental Measurement of the Quantum Metric Tensor and Related Topological Phase Transition with a Superconducting Qubit, 
Phys. Rev. Lett. \textbf{122}, 210401 (2019).
 
\bibitem{Cesium}
T.G. Tiecke(v2.3.2, Sept. 2023),
URL, https://steck.us/alkalidata/cesiumnumbers.pdf.

\bibitem{XJLiu2013}
X.-J. Liu, K. T. Law, T. K. Ng, and Patrick A. Lee,
Detecting Topological Phases in Cold Atoms,
Phys. Rev. Lett. \textbf{111}, 120402 (2013).

\end{thebibliography}
\end{document}